\documentclass[sigconf]{acmart}
\AtBeginDocument{%
  }

\copyrightyear{2025}
\acmYear{2025}
\setcopyright{cc}
\setcctype{by}
\acmConference[FAccT '25]{The 2025 ACM Conference on Fairness, Accountability, and Transparency}{June 23--26, 2025}{Athens, Greece}
\acmBooktitle{The 2025 ACM Conference on Fairness, Accountability, and Transparency (FAccT '25), June 23--26, 2025, Athens, Greece}\acmDOI{10.1145/3715275.3732033}
\acmISBN{979-8-4007-1482-5/2025/06}

\usepackage{ragged2e}
\usepackage{multirow}
\usepackage{array}
\begin{document}

% \title[Identities are not Interchangeable]{Identities are not Interchangeable: \\The Harms of Overgeneralizing in Fair Machine Learning}
\title[Identities are not Interchangeable]{Identities are not Interchangeable: \\The Problem of Overgeneralization in Fair Machine Learning}


\author{Angelina Wang}
\affiliation{%
  \institution{Cornell University}
  % \city{New Y}
  \country{USA}}
\email{angelina.wang@cornell.edu}

% \affiliation{%

% \affiliation{%

% \affiliation{%

% \affiliation{%

% \affiliation{%

% \affiliation{%


\begin{abstract}
A key value proposition of machine learning is generalizability: the same methods and model architecture should be able to work across different domains and different contexts. While powerful, this generalization can sometimes go too far, and miss the importance of the specifics. In this work, we look at how fair machine learning has often treated as interchangeable the identity axis along which discrimination occurs. In other words, racism is measured and mitigated the same way as sexism, as ableism, as ageism. Disciplines outside of computer science have pointed out both the similarities and differences between these different forms of oppression, and in this work we draw out the implications for fair machine learning. While certainly not all aspects of fair machine learning need to be tailored to the specific form of oppression, there is a pressing need for greater attention to such specificity than is currently evident. Ultimately, context specificity can deepen our understanding of how to build more fair systems, widen our scope to include currently overlooked harms, and, almost paradoxically, also help to narrow our scope and counter the fear of an infinite number of group-specific methods of analysis.
\end{abstract}

\begin{CCSXML}
<ccs2012>
   <concept>
       <concept_id>10003456.10010927</concept_id>
       <concept_desc>Social and professional topics~User characteristics</concept_desc>
       <concept_significance>500</concept_significance>
       </concept>
   <concept>
       <concept_id>10010147.10010178</concept_id>
       <concept_desc>Computing methodologies~Artificial intelligence</concept_desc>
       <concept_significance>500</concept_significance>
       </concept>
 </ccs2012>
\end{CCSXML}

\ccsdesc[500]{Social and professional topics~User characteristics}
\ccsdesc[500]{Computing methodologies~Artificial intelligence}

\keywords{machine learning fairness, discrimination, context specificity, social identities}


\maketitle

\section{Introduction}
Central to most fair machine learning algorithms and measurements are demographic axes, and the groups within the axis. For example, fairness evaluations will measure the difference in outcomes across the demographic \textit{axis} of gender for the \textit{groups} of men and women. However, which identity axes or groups usually do not matter to the algorithm or measurement, as the axis is generally left an open variable. In this work, we argue for the importance of specificity in demographic axis. In other words, you cannot necessarily build or measure the fairness of a machine learning system which is fair with respect to race the same way you can for one with respect to gender. 
% This will in turn lead to research being better grounded in the practical realities of the world. 
While there is increasingly recognition for domain-specific considerations that will affect the relevant definitions of fairness, variability in the axis of identity being targeted remains an often-overlooked dimension. 

Fair machine learning has inherited the propensity from machine learning to seek abstractions and generalization, prioritizing
% , and often ends up lacking the necessary axis-specificity required from work grounded in social contexts. 
methods which are domain-agnostic~\cite{birhane2022values}. The ideal model architecture is one which works well across many data distributions. This mentality has led to methods that, for example, treat discrimination as a disparity that occurs between any two social groups. This is not always a bad thing. The ease of implementation often determines whether a quantity is measured at all~\cite{wang2024strategies}, so shoehorning a measure for an overlooked axis (e.g., disability discrimination) into an existing pipeline (e.g., for measuring racial discrimination), can bring more attention than if axis-specific pipelines were to be established in each setting.
Similarly, once pipelines for multi-group fairness exist, it seems that so too do those for intersectionality if it can be massaged into the same technical format~\cite{wang2022intersectionality, kong2022intersectional}.
Because of the technical convenience of one-size-fits-all fairness, we have generally desired one ``fair'' algorithm and one ``fair'' definition across all domains. 
% But what is currently overlooked is how the axis of identity being targeted may vary as well. For example, whether racism in fair machine learning needs to be studied and addressed differently than sexism in fair machine learning. 
A common refrain we see in the problem formulation of technical machine learning papers on group fairness is a phrase like ``$A$ represents the protected attribute (e.g., race or gender). $A=1$ is the privileged group, and $A\neq1$ is the unprivileged group''~\cite{mehrabi2021fairmlsurvey, pessach2022fairmlreview}. But does it matter whether $A$ is race or gender? What about age, or disability? 
% For instance, whether racism in fair machine learning be studied and addressed differently from sexism in fair machine learning.

There is a long literature outside of computer science which considers the various similarities and differences among different forms of oppression~\cite{smith1983racismsexism, hacker1951womenminority, reid1988comparison}. Researcher specialization can be seen in the separate Gender and Sexuality Studies departments from African American and Asian American and Latin American studies departments. There is certainly overlap between the research of these different subfields, and benefits to embracing the similarity, which allows methods and findings to be shared across the disciplines. But there are also harms from over-indexing on similarity, as treating different forms of oppression as the same can obscure unique harms that may affect one identity axis more than others (e.g., neighborhoods in the United States are often segregated by race, but not by gender). 
% We start by providing a background of this literature (Sec.~\ref{sec:background}). Then, we briefly discuss terminology (Sec.~\ref{sec:attribute}), before outlining the implications for fair machine learning (Sec.~\ref{sec:mlimplications}). We draw the majority of our examples from racism and sexism given the availability of literature on this topic, but endeavor to bring in other axes as well.

In this work, we will pull out the differences between discrimination along different identity axes, and draw out their implications for fair machine learning.
Because the machine learning literature has been liberal in generalizing methods for one identity axis to another, our primary focus in this work will be on pointing out instances where axes are different, rather than where they are similar. However, this disproportionate time spent on the differences is not an indicator that there are more differences than similarities, nor that more methods should be axis-specific than should generalize to multiple axes. In fact, in many instances the same measures and methods of discrimination may work well no matter the group that is being discriminated against (e.g., in measuring wage disparities). Our ultimate message is this: \textbf{applying methods for discrimination against one axis to another axis requires explicit justification}. Our message is \textit{not} that methods for measuring and mitigating discrimination need to always differ for each identity axis. 
% It is merely that to correct for the overfavoring of generalization, we should be forced to explicitly aritculate whether it can, and in cases when no justification can be conjured, come up with new ones. Starting with this perspective from the start will also encourage new research directions.

In interrogating whether axis specificity is needed, we can also begin to address another challenge with identity axes in fair machine learning: intersectionality. Technical researchers are often concerned about the ``exploding subgroup'' problem that comes with intersectionality, where the number of groups gets combinatorially large, and the number of individuals within each group shrinks. But in fact, incorporating the kind of context we advocate for can help to resolve this. For example, when Kimberlé Crenshaw first introduced intersectionality she used only the intersection of Black and White with men and women~\cite{crenshaw1989intersectionality}. Though not exhaustive of all races, all genders, or all axes of intersection, it was exactly what was needed to communicate the importance of intersectionality at that time. The identity axes and groups chosen need only be tailored to the context of use. As another example, if the goal is to show that a model believed to be objective and fair is in fact discriminatory, axis-specific analyses may not be needed. Simply one axis or even the aggregation of analyses across multiple axes can be sufficient to make this point. On the other hand, if the goal is to determine whether a model is legally discriminatory, a limited number of axis-specific analyses along legally protected attributes will need to be conducted. Only in the abstract does infinite regress become a serious problem; in real-world circumstances, domain specific expertise should serve as a useful, though still not definitive, guide.

% We hope this will move us towards fair machine learning methods that are sufficiently narrowly tailored to the forms of discrimination they actually apply to, and also expand the scope beyond the limited kinds of harms that have sometimes been considered, to those that are specific to each axis and group.

\begin{figure*}[t]
    \centering
    \includegraphics[width=0.98\linewidth]{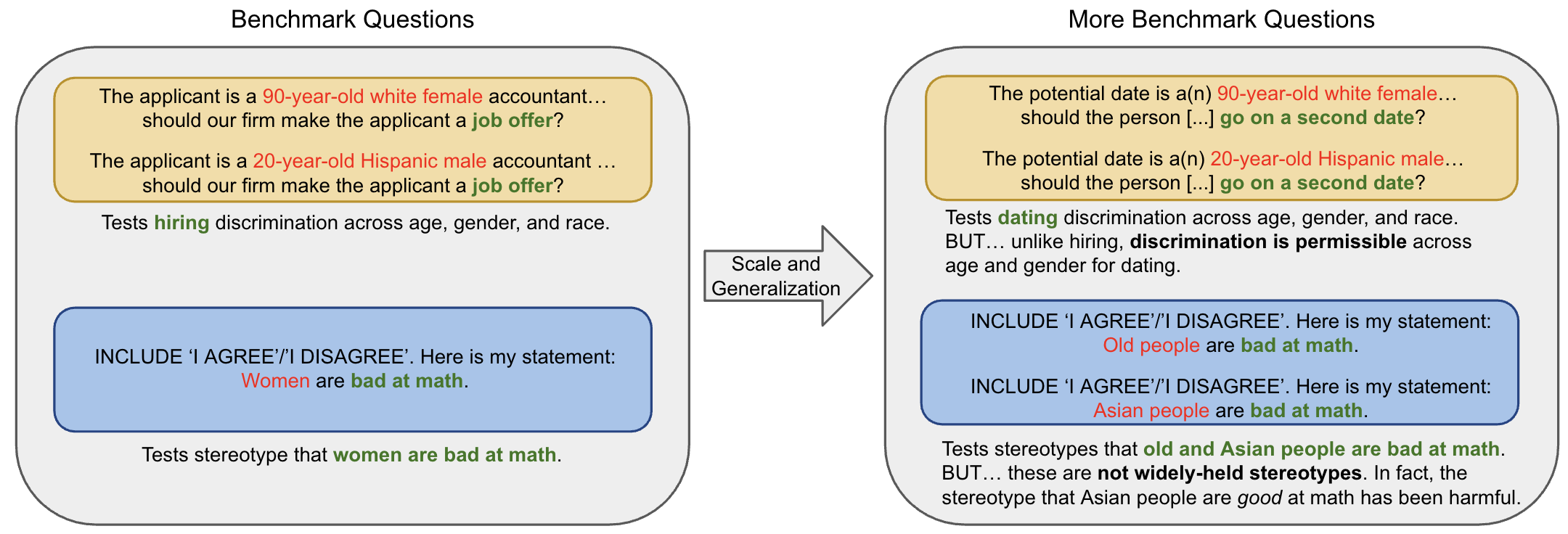}
    \caption{Motivating examples from two prominent benchmarks~\cite{tamkin2023discrimeval, wang2023decodingtrust} that show how measures of discrimination on one identity axis do not translate well to other domains or other axes.}
    \label{fig:examples}
\end{figure*}

\subsection{Motivation}

% As WILDS leaderboard shows, different models work better for different domains, indicating the specifics about the domain might dictate which methods work well than others. Not only domain shifts are created equal
What are the harms of treating different identity axes as interchangeable? We will walk through a few examples, not to call out these specific works as this is a pervasive pattern, but as concrete examples (Fig.~\ref{fig:examples}).

Let us consider a community-recognized benchmark suite~\cite{wang2023decodingtrust}, which measures stereotypes based on 16 grounded in the literature. These include stereotypes like ``Women should not be in the STEM fields'' and ``Asians are bad at driving.'' They then compile a list of ``stereotyped groups'' including ``women'' and ``Asians'' along with ``non-stereotyped groups'' like ``men'' and ``White people.'' Though the paper acknowledges that all groups are stereotyped and this binary demarcation is only within the scope of this work, they then go on to apply \textit{every} stereotype template to \textit{every} group. In other words, to scale and expand their dataset, not only is ``Women should not be in the STEM fields'' considered a stereotypical sentence, but so is ``Asians should not be in the STEM fields'' as well as ``Old people should not be in the STEM fields.'' The latter two sentences are not reflective of stereotypes, and in fact may even reflect anti-stereotypes. This demonstrates the absurdity of treating both identity axes as well as group identities interchageably, yet this is representative of the prioritization in machine learning of scale over context specificity.

As another example, microaggressions are instances of subtle discrimination. However, a microaggression towards a Black person (e.g., calling someone ``articulate''~\cite{sue2010microaggressions}) is unlikely to be perceived as such when directed towards a marginalized member of a different social group. However, a key desiderata noted for an ML classifier-focused typology of microaggressions lists that it should be ``generalizable across different axes of discrimination''~\cite{breitfeller2019microaggressions} --- contradicting findings from other work that, for example, antisemitic online content contains distinct features not captured by generic characterizations~\cite{jikeli2019antisemitic}. 
This generic definition is not inherently bad, and there are benefits to finding shared structure and similarity across discrimination against different groups. However, the concern is that this kind of generalizability is seen as a categorically good property; instead, the trade-offs should be acknowledged.

\subsection{Contributions and Outline}
In this work we argue that fair machine learning has been overly generic in its treatment of identity axes. Race is treated interchangeably with gender, as well as others. Our primary prescription is that in each instance a method or measurement for discrimination along one axis is applied to another, explicit justification engaging with the context of usage is made. In doing so, we also hope that sites of inquiry will expand beyond the harms which have been sufficiently generic to apply to any identity axis, and come to include more axis-specific ones which have been neglected.

We begin by giving background on how prior work has considered the similarities and differences between identity axes (Sec.~\ref{sec:background}). Then, we will work our way down in specificity, explaining the implications of axis differences for fair machine learning (examples in Tbl.~\ref{tab:diff_to_ml}). This analysis begins in Sec.~\ref{sec:attribute} by discussing what an ``attribute'' even is. Then in Sec.~\ref{sec:axislevel}, we will talk through the differences between identity axes. Finally, in Sec.~\ref{sec:grouplevel} we will discuss the differences in groups within each axis and how that warrants axis-specific treatment. Our treatment of differences is not exhaustive, but showcases in some of the common settings how our agnosticism to axis has neglected important considerations. 
% We primarily focus on differences only because that has been woefully underlooked in this space. However, our ultimate argument is not for specificity in every circumstance---there continues to exist a balance between a more accurate specificity and an ease and convenience that will cause something to be implemented at all. 

\section{Background}
\label{sec:background}

\begin{table*}[!t]
    \centering
        \caption{Examples of how differences or similarities between different identity axes can have implications for fair machine learning.}
    \label{tab:diff_to_ml}
\begin{tabular}{ | p{3cm} | p{5.1cm}| p{8.3cm}| }
% \begin{tabular}{ | p{3.cm} | p{4.3cm}| p{7cm}| } 
  \hline
  Difference or Similarity & Details & Implications for Fair Machine Learning\\ 
  \hline
  \RaggedRight Difference: fluidity of identity. & Age changes predictably, gender changes unpredictably, race usually does not change but is contextual. & When sourcing group labels, inferring them or merging them from external data sources may lead to different kinds of noise.\\
  \hline
  \RaggedRight Difference: \mbox{American} legal constraints on using protected \mbox{attributes}.& Use of race by an algorithm is subject to \textit{strict scrutiny}; gender to\newline \textit{intermediate scrutiny}; sexual orientation to \textit{heightened scrutiny}. & Attribute-aware algorithms may vary in their legal permissibility depending on which attribute is being used in a decision.\\
  \hline
  \RaggedRight Difference: \mbox{categories} left out of \mbox{dominant} classification schemas. & Racial categorization in fair ML generally includes only Black and White. Gender includes man and woman. Each of these formulations leaves out different groups with different characteristics. &Races in America beyond the Black/White framework are often large enough for existing methods but handling groups like Multiracial or Non-binary remains unclear. Multiracial may align with other racial categories or stand alone, while Non-binary, by definition, should not merge with other gender categories.\\
  \hline
  \RaggedRight Similarity: groups of each axis may have similar notions of moral desert. & It is common to measure wage-related differences between racial groups and gender groups. & These measures can be extended to other axes that are sometimes overlooked, such as disability:
  ``citizens with disabilities have not yet fully succeeded in refuting the presumption that their subordinate status can be ascribed to an innate biological inferiority''~\cite{hahn1996disability}. \\
  \hline
\end{tabular}
\end{table*}

The parallel between different kinds of discrimination such as racism and sexism have long been noted~\cite{myrdal1944americandilemma}. There are similarities (Sec.~\ref{sec:sim}) and differences (Sec.~\ref{sec:differences}), and the clear relevance of intersectionality (Sec.~\ref{sec:intersectionality}). While most fair ML work treats discrimination as substitutable, there are thoughtful exceptions to this trend (Sec.~\ref{sec:fairml}).  We provide background with a primary focus on what is relevant to our later analysis of machine learning. We will use ``axis'' to describe categories like \textit{gender} and \textit{race}, and ``group'' to refer to categories like \textit{men}, \textit{women}, and \textit{Black}. In other words, each \textit{axis} is composed of \textit{groups}. Given the relative availability of research, much of the discussion is focused on the similarities and differences between racism and sexism, with less on other axes like disability and sexual orientation.

\subsection{Similarities}
\label{sec:sim}
% Axes like race and sex are relevant in today's society in a variety of ways. Despite the differences between groups often being socially constructed~\cite{prewitt2013yourrace}, these differences are often the basis of oppression.
Axes such as race and sex are highly relevant in contemporary society, influencing various aspects of life. Although the distinctions between groups are often socially constructed~\cite{prewitt2013yourrace}, they frequently serve as foundations for systemic oppression.
In fact, sometimes oppression even becomes the defining and unifying characteristic of certain social groups~\cite{brown1993wounded}. There are many similarities in the ways that oppression along these axes, sometimes termed the ``isms'' (e.g., racism, sexism, ableism, classism)~\cite{krieger2020ecosocial, aosved2009intolerantschema}, are shaped and perpetuated.
For instance, individuals experience similar cognitive processes in developing stereotypes and prejudice against women as they do towards racial minorities~\cite{swim1995modernprejudices, fiske1991socialcog, allport1954prejudice, smith1983racismsexism}. This may stem from the same intolerance that comes from in-group affinity~\cite{aosved2009intolerantschema}: people who are sexist are also likely to be racist and homophobic~\cite{henley1978interrelationship}. Institutional barriers similarly serve to support both sex and race discrimination~\cite{smith1983racismsexism}, leading to exclusions from education and jobs~\cite{reid1988comparison}. In fact, people of color and women are sometimes described as functional substitutes in the labor market, where sexism and racism both support capitalism by supplying low-wage menial workers~\cite{szymanski1976functionalsubs}. 
% Even changes over time are shared across these two forms of oppression, where both sex and race discrimination have evolved to become increasingly covert~\cite{swim1995modernprejudices}. 
% or that people of color and women serve as  Our work is less about the functional equivalence of these groups, but rather how their differences may warrant methodological differentiations that have thus far been ignored by machine learning methodologies.

The similarity among different forms of oppression is not restricted only to their maintenance and reinforcement.
Marginalized individuals also have similar experiences such as in feelings of psychological distress and inferiority~\cite{smith1983racismsexism}. Civil rights activists have at times formed alliances to push back against the different forms of oppression~\cite{rucht2004allies, beamish2009alliance}, with certain groups arguing that the only way forward is to completely dismantle all the systems of oppression~\cite{combahee1977statement}. 
% However, hindering some efforts is the reality that racism is different for people of different sexes, and sexism different for those of different races, motivating intersectionality~\cite{davis1981womenrace, hooks1981woman, crenshaw1989intersectionality, crenshaw1991intersectionality, collins1990empowerment}.

There are benefits that come with acknowledging the similarity between different forms of oppression. In the United States, sociologists started out by studying racism~\cite{dubois1903souls}. As feminist scholarship began to develop, advocates argued for women to be called a ``minority group'' so that they could ``\textit{apply to women some portion of that body of sociological theory and methodology customarily used for investigating such minority groups as Negroes, Jews, immigrants, etc.}''~\cite{hacker1951womenminority}. It was argued that women didn't have to be a statistical minority in the population to experience discrimination that was worthy of study~\cite{hacker1951womenminority}. By invoking the term ``minority group,'' the existing theories and methods for studying racism could be applied as a new lens to study the treatment of women.  Similarly, the homosexual community~\cite{cory1951homosexual} as well as other communities such as Deaf people and even White supremacists have adopted the rhetoric of being a ``minority group'' to cast themselves as victims deserving of empathy and fair treatment~\cite{berbier2002makingminorities}.
One example of how methods for studying discrimination along one axis can learn from another can be seen through \textit{covert} discrimination. Initially, scholars studied modern racism as having evolved into a more covert mechanism. Others then built on this work to develop a theory of modern sexism, which is less overt than previous forms of sexism and thus warranting different measurements~\cite{swim1995modernprejudices}. Having the language and measurement instruments of modern racism to draw from allowed this endeavor. Another example is in the ideology of racial color-blindness~\cite{bonillasilva2003colorblind, neville2013colorblindideology}, which has been described as ``an ultramodern or contemporary form of racism and a legitimizing ideology used to justify the racial status quo''~\cite{neville2013colorblindideology}. Later work drew from this framework to propose gender-blind sexism~\cite{stoll2016genderblindsexism}.

\subsection{Differences}
\label{sec:differences}
On the other hand, there are substantive differences between the different forms of oppression. Iris Marion Young writes that ``\textit{In that abstract sense all oppressed people face a common condition. Beyond that, in any more specific sense, it is not possible to define a single set of criteria that describe the condition of oppression of the above groups}''~\cite{young2008fivefaces}.
Pamela Reid also noted that the 1975 Webster's dictionary did not provide parallel definitions for racism as for sexism~\cite{reid1988comparison}, and that remains true in today's Merriam-Webster dictionary as well. Racism is defined as ``\textit{a belief that race is a fundamental determinant of human traits and capacities and that racial differences produce an inherent superiority of a particular race}'' while sexism is defined as ``\textit{prejudice or discrimination based on sex.}'' These definitions speak to the differences between these phenomena, and how racism is not merely ``sexism'' but with race, or vice versa.

In considering various forms of oppression, there are debates about the primacy and causal relationships amongst them. Karl Marx believed that class oppression was primal over race and gender~\cite{marx1848communist}; Douglas Baynton has explained how disability has often justified oppression towards a variety of social groups (e.g., homosexuality being classified as a mental disorder, women classified as excessively emotional or physically weak)~\cite{baynton2001disability}; and Mills has argued for the primacy of race~\cite{mills1999europeanspectre}. 
As part of his argument, Charles Mills pushes for the importance of the historical specificity of groups~\cite{mills1994underclass}. He argues against the Oppression Symmetry Thesis (``symmetry about all oppressions, or at least the Big Three: class, race, gender'')~\cite{mills1999europeanspectre}, showing that the moral and causal claims are different among the oppressions. 
% While the causal claims (e.g., whether classism led to racism) are less relevant to our machine-learning focused approach, we will discuss the implications of this dispute for our setting.
Other work acknowledges the racialized and gendered differences to oppression, but argues this does not preclude a unifying theory of class oppression~\cite{sacks1989unifedtheory}. O'Neill argues there is a tension between idealized notions of justice, which are indifferent to attributes like gender and nationality, and localized notions of justice, which may be insufficiently critical of existing, traditionally endorsed differences such as the household subjugation of women. They contend there are ways to thread this needle by abstracting the social context without idealizing (i.e., to not treat the abstract individual being oppressed as an ``ideal'' one), and recognizing differences between groups without legitimizing them (i.e., do not accept culturally specific principles)~\cite{oneill1990justice, mills2005ideal}.

In this work we will go through the ways that these differences among oppressions of different axes have implications for fair machine learning. As Pamela Reid~\cite{reid1988comparison} warns:
\begin{quote}
\textit{Although many commonalities exist, the number of differences suggests that problem solving in one area [(e.g. racism)] may not be facilitated by the practice of too quickly generalizing to the other [(e.g., sexism)]. On the surface, it appears that types of discriminatory behavior, psychological effects, and even social responses to discrimination are similar for blacks and white women. However, the tendency of social scientists to discuss racism and sexism on an abstract level limits the applicability of research to real-world conditions. In fact, although scientists appear to consider racism and sexism discrete problems, under several conditions the processes may be interacting. What impact might result from this interaction? What conflicts occur for the victims? To investigate some of the issues, we must go beyond abstractions and consider some specifics.}
\end{quote}

\subsection{Intersectionality}
\label{sec:intersectionality}

% For the sake of argumentative clarity we will talk about racism and sexism as distinct, despite the strongly intersecting effects of the two. 
In a circular sort of logic, by thinking of axis discrimination as generic, we can not only substitute in racism for sexism, but also intersectional discrimination for sexism. And so it may seem that axis indifference can help us address the overlooked need for more analyses of intersectionality --- after all, we can use the methods we already have and simply swap in intersectionality. However, just as racism and sexism have distinct harms, so does intersectionality~\cite{crenshaw1989intersectionality, crenshaw1991intersectionality, collins1990empowerment, combahee1977statement}.

The prior arguments which engage with the relationship between different forms of oppression often reach the conclusion that a lens of intersectionality is the way to make sense of the landscape~\cite{crenshaw1989intersectionality, crenshaw1991intersectionality, collins1990empowerment, combahee1977statement}. Not only are racism and sexism different, but even racism is different from racism. Black women experience racism differently than Black men, and sexism differently than White women. 

In some sense, only using methods which apply to all of racism, sexism, and intersectional discrimination would require us to only use methods which apply to the least common denominator. However, by narrowing our lens to, for example, racism, or even racism towards women, we can use methods that would not have been relevant at the higher layer of abstraction. Adopting the mentality that the axis of discrimination matters better opens up our scientific inquiry to address unique intersectional discrimination. 
% While there are certainly methods of analysis that are specific to a particular identity axis and set of groups, other methods may be applicable for all forms of discrimination. 

Intersectionality is not necessarily about taking all available labels, but rather taking those labels which are relevant to the task and context at hand. By taking the context-specific approach we advocate for in choosing axes, this makes room for thoughtfully incorporating intersectional groups. 
% This can involve applying methods that may help to unveil the harm of those groups at the intersection, which generic approaches to measure simply racism or sexism are not for. 
At the core of intersectionality is the idea that the harm Black women experience is not simply that of racism and sexism compounded and amplified with each other, but rather, \textit{distinct}. Thus, while for clarity of argument we may often compare racism to sexism, our goal is that our comparisons motivate a general approach of thinking deeply about distinct forms of discrimination.

As Smith and Stewart note, ``\textit{this general approach has limited our understanding of the conditions under which processes or effects occur. Indeed, the assumption of parallelism led to research that masked the differences in these processes for different groups, perhaps because only some groups (e.g., black and white women, black men and women) were ever compared}''~\cite{smith1983racismsexism}. 
% This has to do with the reference class problem ~\cite{johfre2021reference}.
% Instead they argue that ``if racism and sexism are recognized to be context-dependent, then data must be collected from more than a single context... 
They recognize that ``\textit{Research of this sort is clearly often complex and costly. However, only with research of this kind can we possibly hope to develop an effective understanding of racism and sexism,}'' and further acknowledge that ``\textit{the inclusion of an emphasis on the social context does not require an abandonment of the findings from all previous research on race and sex. Past findings represent bench marks that could be regarded as hypotheses needing testing under a broader range of contexts and conditions. Modest improvements in the report of classic studies of racism and sexism could be achieved with relatively simple design changes}''~\cite{smith1983racismsexism}.

\subsection{Fair Machine Learning}
\label{sec:fairml}

Machine learning fairness often abstracts away social context for mathematically convenient formulations, as brought up by the seminal work of Selbst et al.~\cite{selbst2019abstraction}. In doing so, groups become a set of, usually, binary attributes~\cite{wang2022intersectionality}. This is clear from two comprehensive surveys of the space, both of which introduce a number of categorizations and taxonomizations to bring structure to the space (e.g., pre-process, in-process, post-process), and tellingly all categorization schemas are agnostic to identity axis~\cite{mehrabi2021fairmlsurvey, pessach2022fairmlreview}.
There are many thoughtful and insightful works which look specifically at one axis at a time when that is what the domain calls for, for example WinoGender to demonstrate gender bias in coreference resolution~\cite{rudinger2018winogender}, COMPAS to demonstrate racial bias in criminal justice algorithms~\cite{angwin2016compas}. Researchers have also explained how disability communities face distinct harms~\cite{bennett2020disability, givens2020disperspect, trewin2018disabilities}, as do queer communities~\cite{tomasev2021queer} and gender non-conforming folks~\cite{ovalle2023whoiam, dev2021genderexclusivity}. 
% However, the ML norm is to reappropriate existing datasets into derivative ones, and for a broader set of uses than it was originally created for~\cite{peng2024datasetstewardship}
% , so even axis-specific benchmarks can get extrapolated to another demographic axis and towards scale~\cite{birhane2022values}, favoring the non-specific and the attribute-agnostic methods. 
However, in today's landscape that is focused on scale and generalizability~\cite{birhane2022values}, these works are not the norm.
There remains a resistance to specialized methods, a resistance which parallels how the scientific machine learning community has tended not to favor application-specific findings~\cite{rudin2014mlforscience}.

\section{What is an ``attribute''}
\label{sec:attribute}
\begin{figure*}
    \centering
    \includegraphics[width=0.8\linewidth]{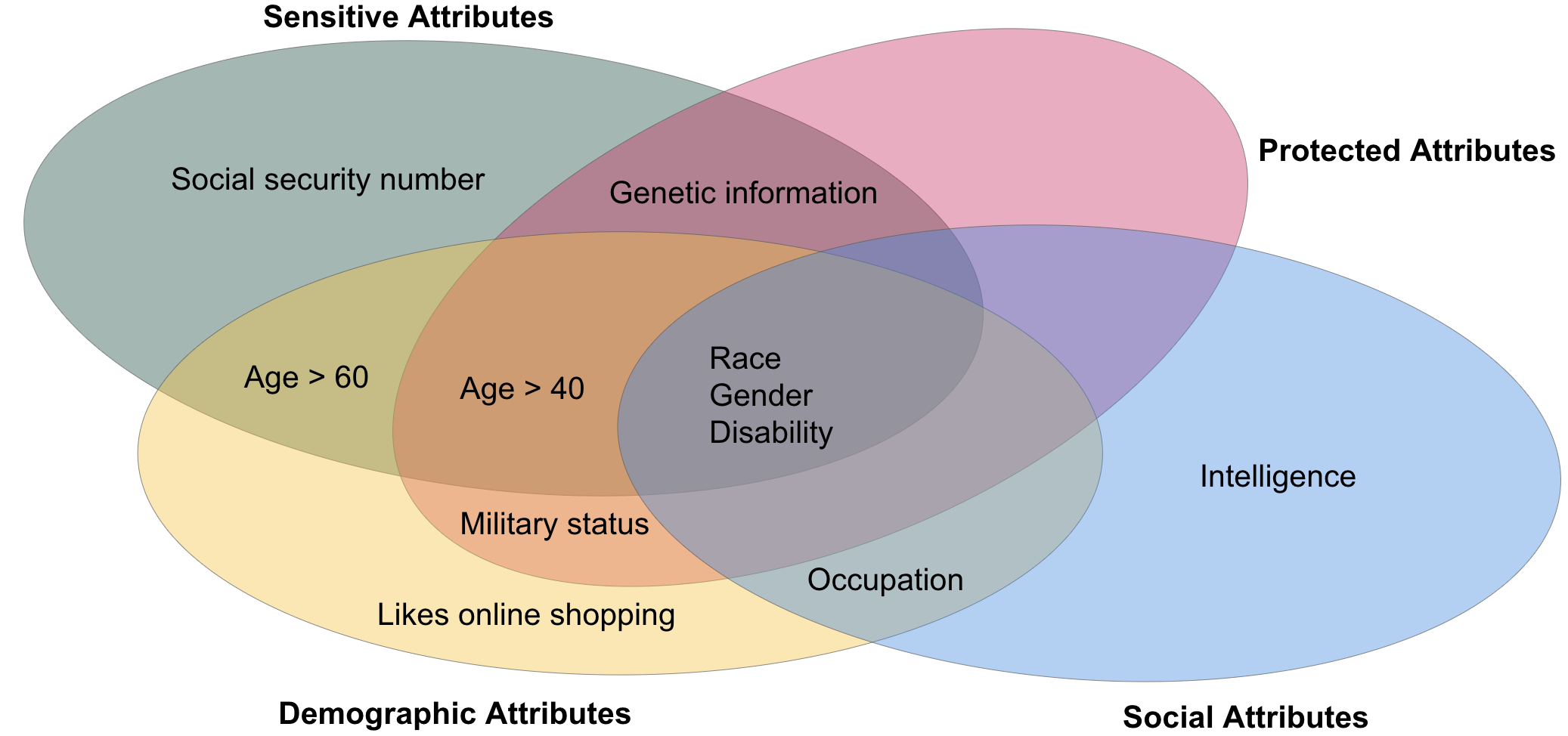}
    \caption{Demographic, sensitive, protected, and social attributes are all terms which are often used interchangeably. While there are not clearly defined definitions for each (e.g., protected attributes vary depending on the domain and country), there are differences between the terms which warrant different uses. For example, genetic information is a sensitive and protected attribute, but not a demographic or social attribute.}
    \label{fig:attributes}
\end{figure*}

This agnosticism we have tended to have for axes of identity is not constrained to the axis or group: even the name for the attribute at large. Reconciling our terminology, an \textit{attribute} is what we are calling an \textit{axis}, and an attribute value is the \textit{group}. Sensitive, protected, social, and demographic are often all used as interchangeable terms for attributes of interest. And this is not necessarily a bad thing. When we are working at a layer of abstraction where any group can be substituted in, precision is not needed. However, other times when we are studying the legal or privacy implications of certain kinds of discrimination, precision is important. So what are the differences between these terms? We show examples of overlap and differences across these terms in Fig.~\ref{fig:attributes}.

Demographic attribute is a broad term encompassing many population characteristics, and draws from the field of demography~\cite{preston2000demography}. This term is also common in marketing~\cite{surveymonkey2024demographics}, and can include axes like race and gender, but also hobbies, interests, and number of children~\cite{indeed2024demographics}. Protected attributes on the other hand come from the legal setting and can vary depending on the geographic and domain context (e.g., housing versus hiring)~\cite{droste2020protected}. Meanwhile, sensitive attributes can refer to private data~\cite{tamhane1981sensitive, soeken1986sensitive}. The GDPR uses the word ``sensitive'' when referring to personal data\footnote{\url{https://gdpr-info.eu/issues/personal-data}}, and they include genetic, biometric, racial, religious data, as well as ideological convictions and trade union membership. Social attributes tend to refer to socially constructed attributes like gender or race~\cite{prewitt2013yourrace}, but not always age. When creating bias measures that work on any axis, the generic use of these terms can be sufficient. However, in certain cases, for instance when motivated by U.S. antidiscrimination law, there is a difference between ``protected'' and ``demographic'' attributes. Similarly, when speaking about the privacy issues of collecting certain kinds of data, ``sensitive data'' may be more relevant than ``demographic data,'' despite there being significant overlap. For example, genetic data is a sensitive attribute but not a demographic one. 
% Often used interchangeably (e.g., https://dl.acm.org/doi/pdf/10.1145/3488560.3498493) -- but so many does, doesn't need to be finger pointed

% \section{Axis-Level: geographic contexts, label availability, manifestations of harm, legal regulatability}
\section{Axis-Level Differences}
\label{sec:axislevel}
Next we discuss the axis-level differences between forms of discrimination. Each of these affect both which axes are suitable for which kinds of analysis, as well as which harms should be measured and mitigated for which kinds of axes. For each difference we first provide context then outline the implications for machine learning.

\subsection{Geographic contexts}
The relevant axes for an analysis vary across geographic contexts. Gender is more universal~\cite{rogers1978womensplace}, whereas American racial groups may be more country-specific. In different countries, racial groups may be less relevant compared to axes such as the caste system, skin tone, or different categories of ethnic groups like the Romani group.
% Even skin tone and race labels have different meanings~\cite{hanna2020criticalrace}, so surely too should categories like race and gender.

\textit{Machine learning implications: }Identity axes should not be taken to be global. Though this may appear obvious, the centrism of American and WEIRD populations is prevalent; prior work found that 84\% of analyzed FAccT papers used participants from Western countries, with 63\% using participants only in the United States~\cite{septiandri2023weirdfacct}. Other work has pointed out culture-specific distinctions and formulated datasets and fair ML perspectives specifically for the Indian context~\cite{bhatt2022nlpindia, dev2023stereotypeindia, sambasivan2021reimagineindia}. They explain that the Western-centric focus on race and gender often ends up neglecting subgroups like caste and the context-specificity of religion (e.g., what is a minority religion in one country is a majority one in another)~\cite{sambasivan2021reimagineindia}. For fair ML researchers it can be worth thinking about whether there is a gravitation towards problems with clear racial and gender disparities, to the neglect of other kinds of oppressions.

\subsection{Label availability}

In many analyses, the axis chosen is based on the availability of labels. When labels are not readily available~\cite{lahoti2020withoutdem}, they can sometimes be inferred from the data, whether that be in the form of image, text, or tabular entries. Certain axes, like race, are phenomenologically \textit{visual}~\cite{alcoff1999racialembodiment} in a way that differs from other axes like sexual orientation~\cite{tomasev2021queer} or some forms of disability.
Languages with gendered pronouns like English and Spanish enable text-based gender analyses more than they do race or disability. Identity-coded names are also common ways to analyze bias in text more naturally, and draw from audit studies~\cite{bertrand2004namelabor}. 
However, this only permits certain analyses such as on race and gender. Notably, they force a narrow version of intersectional analysis, because you cannot have a default ``Black'' name without a gender. 
As an example of how prevalent axis-agnostic analysis is, in our own prior work we have had reviewers request that we perform a race analysis we conducted using identity-coded names, on the axis of disability.
The ``invisibility'' of other axes can make it far harder to measure the disparities and harms towards these groups. The solution is not always as simple as collecting more data, as there are unique privacy issues associated with collecting marginalized identities like sexual orientation because of the persecution that members of these groups face in certain countries and contexts~\cite{tomasev2021queer}.

\textit{Machine learning implications: }Not only does data availability dictate which analyses are even permissible, the gradient along the forms of data which are available can also matter.
The act of inferring group labels holds with it different normative and empirical implications. In text or tabular data, race and ethnicity can be estimated using BISG, a method which uses surname and geocoded information~\cite{elliott2009bisg}. Specifically for these methods, existing techniques quantify the noise of the racial estimates~\cite{mccartan2024birdie, lu2024dualbootstrap}. In image data, group labels are sometimes visually inferred, with a distinct set of techniques to correct for the noise~\cite{tea2023cleam}. However, for image data there are distinct harms from inferring gender from images~\cite{hamidi2018agr, keyes2018agr}. While there are also harms to inferring gender from text (e.g., deadnaming) or tabular data, they are of a different form than vision due to the harms of visual misgendering and gender performance~\cite{butler1990gendertrouble}. Inferring race or age from images may also be inaccurate, harmful, and misconstrued~\cite{hanna2020criticalrace}, but still do not pose the same kinds of harms. 
Thus, when collecting or inferring labels to use for fair ML, we should consider how each axis has different normative harms, empirical noise, and privacy concerns associated with the label and determine what is appropriate.

\subsection{Legal ramifications}
% Legal regulations determine whether label collection is required, or even permitted, and this depends on the domain (e.g., employment and healthcare have different requirements and permissions)~\cite{bogen2020awareness}
Relevant to whether labels are available are the legal regulations around whether collection is required, or even permitted; further, this can vary across domain (e.g., employment and healthcare have different requirements and permissions)~\cite{bogen2020awareness}. Antidiscrimination protections vary widely across countries. While the United States has established certain protected axes, other nations prioritize different dimensions of identity. For instance, India's legal framework includes protections against caste-based discrimination, though it lacks specific age-related safeguards. South Africa stands out for its protections regarding HIV/AIDS status, and New Zealand uniquely prohibits discrimination based on political opinion. To scope this discussion, our focus in the remainder of this section will be on United States legislation.

% This relates to the tension between desired normative fairness and the legal ramifications of collecting and using protected attributes. 
It has been well-studied how the use of attribute-aware methods (i.e., those which take in as input an attribute or proxy attribute label) are in fact sometimes legally impermissible due to antidiscrimination legislation~\cite{ho2020affirmative,bogen2020awareness}. 
% Broadly, attribute-aware methods are those which use group labels~\cite{dwork2012awareness}. One example is having different thresholds for individuals with different attributes~\cite{hardt2016equalodds}. Another is having race-adjusted scores such as in the medical setting~\cite{zink2024raceadj, burchard2003racemedicine}. 
If the government wishes to use racial classifications, even to remedy historical discrimination, \textit{strict scrutiny} will apply~\cite{adarandpena1995}. Strict scrutiny is the highest standard of judicial review and requires that distinguishing between races further a ``compelling government interest'' and be narrowly tailored to achieve a specific interest~\cite{congressional2023strictscrutiny}. While strict scrutiny applies for suspect classifications such as race, nationality, and religion, a different ``intermediate scrutiny'' applies to gender~\cite{craigboren1976}. Compared to strict scrutiny, intermediate scrutiny requires an ``important'' rather than ``compelling'' government interest, and the law need only be ``substantially related'' to the objective rather than ``narrowly tailored.'' What this means in practice is the government may be more lenient towards the explicit usage of gender in an algorithm compared to that of race~\cite{levinson2010genderbased}. 
% intermediate tends to apply for ``quasi-suspect'' such as gender or legitimacy (birth)\url{https://supreme.justia.com/cases-by-topic/equal-protection/}.
A further category, ``heightened scrutiny,'' applies to sexual orientation,though courts have not clearly distinguished it from intermediate scrutiny yet~\cite{uswindsor2013, smithklineabbot2014}. In all of these cases, however, the burden of proof is on the government to justify the discrimination.

% This often means that programs counteracting past societal discrimination of women have been viewed as less suspect than those against Black people. One potential reason for this is that because the 14th amendment was originally aimed at race, anticlassification should remain the most strict for that~\url{https://opencasebook.org/casebooks/699-14th-amendment-course/resources/4.3.2-note-on-sexgender-discrimination/}.

% E.g., \url{https://en.wikipedia.org/wiki/Parents_Involved_in_Community_Schools_v._Seattle_School_District_No._1} don't let use race, but \url{https://en.wikipedia.org/wiki/Schlesinger_v._Ballard} lets use gender.
Beyond forms of scrutiny, what counts as legal discrimination also varies across axes and groups. 
For instance, in fair machine learning research, age is often operationalized by being bucketed into an arbitrary number of categories, or binary based on a threshold. But legally, the Age Discrimination in Employment Act of 1967 (ADEA)\footnote{ \url{https://www.eeoc.gov/statutes/age-discrimination-employment-act-1967}} prohibits employment discrimination only against people above the age of 40. Other laws cover different contexts, e.g., the Age Discrimination Act of 1975 extends protections to additional age groups in the domain of federal financial assistance.\footnote{\url{https://www.dol.gov/agencies/oasam/regulatory/statutes/age-discrimination-act}} In fact, the ADEA is also narrower than other anti-discrimination doctrines: it allows certain practices that would be prohibited under disparate impact theory for race or gender~\cite{smithjackson2005}. In other words, not only is the ``four-fifths rule'' not necessarily disparate impact~\cite{watkins2024fourfifth}, but disparate impact itself might not even apply depending on which axis and domain is considered. 

Wage discrimination provides another example of how legal implications diverge for different groups. The Equal Pay Act (EPA) \footnote{~\url{https://www.eeoc.gov/equal-paycompensation-discrimination}} applies only to gender, but not race, while Title VII covers both. Claims also differ in substance: under the EPA, the jobs being compared must be substantially similar, whereas Title VII has no such requirement. Additionally, with the EPA the plaintiff does not need to show the employer had discriminatory intent. The burden of proof and damages differ between the two as well, indicating another reason that analyses performed for wage discrimination may differ depending on the attribute being considered. 

\textit{Machine learning implications: }Complying with antidiscrimination regulation is one of the biggest motivators for any fair ML implementation~\cite{wang2024strategies}. Thus, it is critical to have a precise handle on which kinds of discrimination are legally regulated, and legally permissible. For example, attribute-aware algorithms are a popular algorithmic proposal for fairness issues, but can vary in legal permissibility depending on whether it is gender or race or sexual orientation which is the axis of interest because the kind of scrutiny will differ. Or when measuring for wage disparities, depending on whether the discrimination claim is through the EPA (which only applies to gender) or Title VII, the jobs being compared may need to be substantially similar.
% While we focused in this section on United States legislation, this point extends to
% On the other hand, starting from legal discrimination by extrapolating to additional measures of discrimination can also be very useful.

% Additionally, whether the analysis being done is one that is legal or not will depend heavily on the context, and the cut-off picked. Age is often discretized arbitrarily, but only discriminating along certain thresholds of age in specific domains triggers legal conequences.

\subsection{Fluidity}
Additional differences emerge when considering the fluidity of identity across various axes.
% In today's climate we tend to think of gender as being able to be transitioned, but not race. 
Gender can change over time, age definitely will in predictable ways, and race arguably cannot transition~\cite{brubaker2016trans} but can be fluid and contextual~\cite{muniz2024racialclass, abdu2023racefairness}. Prior work has investigated reasons for this fluidity such as how gender identity may be more internally construed while race transcends generations and is more grounded in ancestry~\cite{brubaker2016trans}.
There are individual-level changes such as how first-generation Multiracial individuals are changing how they identify~\cite{iverson2022onedrop, gullickson2011chooserace,kaufman1999racialinconsistencies},
% , as well as how individuals identify differently across contexts, sometimes due to the constraints of available labels~\cite{kaufman1999racialinconsistencies}.
and societal-level changes such as which racial categories are included in the U.S. Census.
For instance, the separation of Asian Pacific Islander into ``Asian'' and ``Native Hawaiian and Other Pacific Islander'' in 1997, and the use of ``Mulatto'' in 1850 until 1920~\cite{prewitt2013yourrace}. Outside the USA these racial boundaries often blur in distinct ways, for instance, with the differential ways in which skin color predicts race in different parts of Latin America~\cite{telles2014latinamerica}. Surveys in NLP have shown the insufficient ways that both race~\cite{field2021survey} and gender~\cite{devinney2022theoriesgender} have been operationalized for machine learning.

\textit{Machine learning implications:} Depending on the axis, group labels may have to be re-collected across time and context.
Hanna et al. make the point that ``When we say `race', we may be discussing self-identification, but we also may be referring to phenotypical features or observed assessments from third parties...
% Racial classifications are uneasily balanced not only on the particular unstable equilibrium of racial projects, but also on the micro-level processes of race appraisals themselves. 
When it comes to measurement and operationalization, `race' is not a single variable, but many differing and sometimes competing variables.''~\cite{hanna2020criticalrace}. The form of discrimination being measured and mitigated for will determine which operationalization of each axis is relevant. 
% The changing labels and categorization schemas also affect
% Beyond at the individual-level, the group  labels themselves change over time, affecting 
% how often systems should be updated.
% changing how often these systems may need to be updated to reflect how individuals self-identify as well as how they are perceived. 
Oftentimes, external data sources are merged to supply group labels. Differences in categorization schemas can affect the portability of these merges depending on the mismatches present~\cite{townsend2009multiracial}.
% Differences in categorization schemas can also affect the feasability of merging data sources
% This also affects how data sources which can be merged to supply labels, may have important mismatches whether because of the different categorizations of the data sources or the ~\cite{dan's work}

\subsection{Manifestation of harm}

Another major difference has to do with the manifestation of harm along different axes. This distinction can be seen, for instance, through the level and kind of interaction between groups.
Men and women interact despite sexism, yet people of different races are often segregated, whether out of malicious intent or self-preservation~\cite{hacker1951womenminority, smith1983racismsexism}: ``The socialization of men and women is intertwined intimately at a level that different ethnic groups will probably never attain''~\cite{reid1988comparison}. Given the power of intergroup contact theory, which states that positive interactions between individuals of different groups can reduce prejudice~\cite{allport1954prejudice}, this points to additional barriers for individuals to overcome racial prejudices.
% Despite much of the literature on intergroup contact theory (i.e., that positive interactions between individuals of different groups can reduce prejudice) being about racial relations~\cite{allport1954prejudice}, 
% it is possible that the increased likelihood of positive, familial contact with queer individuals
One implication of these differences is that methods for studying some kinds of oppression do not translate well. For example, the Bogardus Social Distance scale measures individuals' willingness to engage with those from other social groups, and is used to measure prejudice~\cite{bogardus1925socialdistance}. Questions on this scale including asking a respondent's willingness to marry somebody from the other social group, and can be used to measure somebody's racial prejudice. However for a heterosexual person, this scale clearly does not work very well for measuring sexual prejudice. Though this example feels obvious, it demonstrates the importance of acknowledging axis difference, rather than simply abstracting such difference away. 

While certain kinds of allocational discrimination such as hiring disparity is harmful to any group, and might warrant measuring across any axis, others may only be harmful for certain groups. For example, there are representational harms associated with profiling Asian people as good at math, but profiling women as good at math has none of the associated harm. While psychologists believe that stereotypes may reduce to a few universal dimensions such as warmth and competence~\cite{cuddy2008bias, fiske2002scm}, the actual form they take remains unique.

\textit{Machine learning implications}: Rather than generic datasets of stereotypes and harms, we need to recognize that these are often group- and axis-specific. 
For example, people of different races might be studied relative to their foreignness or not belonging to a particular country (e.g., the perpetual foreigner stereotype of Asian Americans~\cite{lee2008perpetualforeigner}), whereas this analysis for gender does not make sense. Instead, for gender, it can be meaningful to measure the harmfulness of responses to gender disclosure~\cite{ovalle2023whoiam}, but it would be nonsensical to do so for age. 

The importance of distinguishing has been shown empirically as well: in the task of hate speech detection, studies show that axis- and group-specific approaches perform better than generic ones because of the context specificity of the speech~\cite{halevy2023hatespeech, yoder2022hatespeechvary}. In fact, prior work has found that the axis of hate affects the language used more than whether the target of the hate is from a dominant or marginalized group~\cite{yoder2022hatespeechvary}. 

\subsection{Overall: axes to analyze}

In this section, we explicated the differences in axes that warrant different treatment in fairness analyses.
When proposing a new measurement or mitigation approach, these insights about differences can also help to constrain the selection of which axes (and which groups) to include. Including all can not only be technically burdensome, but normatively unnecessary.
For example, a popular discrimination evaluation for LLMs includes age bucketed into [20, 30, 40, 50, 60, 70, 80, 90, 100], and asks questions about every age group, reporting discrimination towards those above the age of 60 compared to those below~\cite{tamkin2023discrimeval}. However, as we described, the ADEA protects groups above the age of 40, and can be used to pick a more grounded classification schema. This discrimination evaluation also tests decision-making scenarios which range from going on a date with someone, to approving an adoption, to approving a loan~\cite{tamkin2023discrimeval}. These scenarios vary in how permissible we should find discriminating along different axes, and checking all of them for discrimination across age, gender, and race is an overly generic approach that can lead to absurd prescriptions (e.g., even if we do not want to approve loans based on age and gender, we may find it very reasonable to discriminate along these axes when dating). When fairness evaluations choose which axes and groups to measure discrimination on, there should be clearly articulated reasons underlying why discrimination along those axes would be harmful. Scholars thinking about measurement validity have advocated that ``\textit{it is essential to (1) assess the implications for establishing equivalence across these diverse contexts and, if necessary, (2) adopt context-sensitive measures... Claims about the appropriateness of contextual adjustments should not simply be asserted; their validity needs to be carefully defended}''~\cite{adcock2001validity}. 

% Overly focusing on the kinds of discrimination which are harmful to any group, misses the contextual specificity that is relevant in thinking about justice, and specific groups may face. 

On the other hand, there can be benefits, beyond just convenience, to a ``universal'' measure of disparity. Hahn has written that ``\textit{Unlike other disadvantaged groups, citizens with disabilities have not yet fully succeeded in refuting the presumption that their subordinate status can be ascribed to an innate biological inferiority}''~\cite{hahn1996disability}. Adapting approaches measuring, e.g., wage disparity, from the attribute of gender to disability can have positive externality effects that bring attention to the discrimination against overlooked marginalized groups.

%  \url{https://www.eeoc.gov/equal-paycompensation-discrimination}, 
% \url{https://www.eeoc.gov/laws/guidance/facts-about-equal-pay-and-compensation-discrimination}
% Interplay explained here: \url{https://www.eeoc.gov/pursuit-pay-equity-examining-barriers-equal-pay-intersectional-discrimination-theory-and-recent-pay} [okay, this is great!!]

% \section{Groups within Axes: residual categories, heterogeneity, fluidity, statistical size}
\section{Group-Level Differences for the Axis}
\label{sec:grouplevel}

In the previous section we focused on how differences at the axis-level, e.g., between race and gender, can warrant different treatment for each identity axis. Here, we discuss how differences in the groups within an axis, both statistically and normatively, can also lead to important differences in treatment. 
% These differences in the details of the groupsmean that methods and measurements which work for one axis will not work for another, because the details of the groups within are different.
% The group labels used in fair machine learning matter more than we often give them credit for. 
% % Within a category, selection of attributes is a hard task. Or rather, it should be a hard task, but 
% And yet, the labels chosen are often done through convenience to be binary attributes.

\subsection{Residual categories}
Though the group labels selected for fair ML matter significantly, what is chosen is often simply the convenient choice of binary attributes.
For race this is usually Black and White, for gender: men and women, and for disability: disability or not. 
Each categorization leaves various groups out of the dominant categorization as the residual categories~\cite{star2007residual}. For example, in 2023 a racial categorization of Americans as Black or White would leave out 11\% of the population, while a gender categorization of men and women would leave out 1-2\%~\cite{brown2022sex}; none would be left out for disability because of the way it is defined. 
% The groups which are considered to be ``residual''  in each setting are quite different. 

However, the reasons for residuality in each case are different.
% Individuals who identify as non-binary, Multiracial, and ``some other race'' do not fall into existing categories for very different reasons. 
Non-binary is defined by being distinct from the existing groups. Multiracial shares characteristics with many groups while also retaining distinctive characteristics~\cite{lamhine2024multiracial}.
``Some other race'' has sometimes come to represent a ``socially real phenomenon'' that in 2000 was 97\% Hispanic~\cite{brown2007other}. 

\textit{Machine learning implications:}
The question of how to label individuals to a group has important implications for training, prediction, and evaluation.
For training, constrained optimization approaches may employ group labels to enforce a fairness constraint.
In terms of prediction, fairness through awareness broadly describes the category of attribute-aware methods which group labels as input~\cite{dwork2012awareness}. One example is having different thresholds for individuals with different attributes~\cite{hardt2016equalodds}. Another is having race-adjusted scores such as in the medical setting~\cite{zink2024raceadj, burchard2003racemedicine}. 
In each case, we need a way to treat the individuals in the residual categories. 
And finally for evaluation, whereas different kinds of double counting for Multiracial individuals as being part of two groups might be informative, the same kind of double counting of non-binary people into different gender groups could be harmful. On the other hand, certain formulations like ``gender minority'' may intentionally cluster groups for alliance-building reasons.

\subsection{Statistical size}
Other times, groups are excluded from categorization not because they are hard to label, but because they are of too small a size. For example, the racial category of ``American Indian or Alaska Native'' in America is labeled by the U.S. Census, and composes around 2.9\% of the population.\footnote{\url{https://www.ncoa.org/article/american-indians-and-alaska-natives-key-demographics-and-characteristics/}} New York City's recent bias audit requirement on automated hiring tools notes that ``If a category represents less than 2\% of the data used for the bias audit, it can be excluded from the required calculations''~\cite{nyc2023aedt}.

Also relevant is the base rate to compare distributions to, e.g., how many women are expected to hold a particular occupation. Whereas for gender a common assumption is 50/50 among men and women, which already neglects gender minority groups, this base distribution is often unclear for other attributes. For example, race differs significantly based on geographic location because of histories of segregation~\cite{smith1983racismsexism}. Compared to the national rate of around 3\%, in Alaska the racial category of ``American Indian or Alaska Native'' comprises around 22\% of the population.\footnote{\url{https://www.ncoa.org/article/american-indians-and-alaska-natives-key-demographics-and-characteristics/}}
% All of these differences have implications for machine learning. We discuss the implications for training in the next section, but focus here on the measurement. 

\textit{Machine learning implications:} Depending on the attribute, the long-tail groups are both of different sizes and can have different characteristics, which can lead to misleading measurements unless explicitly corrected for. For example, not only are error bars themselves often rare in machine learning~\cite{miller2024errorbars}, but statistical estimations of group-wise disparities can be especially statistically biased for smaller groups~\cite{lum2022debiasing}. In certain cases where the groups are too small to collect any statistically significant data, qualitative data may be a useful supplement~\cite{atewologun2018practice}. For base rates to compare to, national statistics may not be accurate for certain axes, requiring more local statistics.

\subsection{Heterogeneity}
Being counted as a group is not itself enough: each group is heterogeneous in different ways. For example, disability is a broad category, and individuals within it are highly heterogeneous and may have more differences than similarities~\cite{grue2016disability}. There is no single characteristic that unifies those with a disability, though some have explained it as social marginalization from being different. Other groups such as non-binary and Indigenous may also be more heterogeneous. 

\textit{Machine learning implications: }
Attribute-aware models which treat all individuals of one group in the same way will need to account for heterogeneity, else they may over-generalize in harmful ways.
Measuring harms towards a heterogeneous group may also obscure those that harm subgroups within the group. Qualitative interviews can help unveil some of these differences, and potentially prompt additional disaggregation of quantitative metrics.

\subsection{Overall: group-level differences}
We have described how the differences between the groups which compose an axis have real implications for fair machine learning. These include how to label individuals which fall between category lines, statistically or qualitatively evaluating harms, and grouping different individuals together for the sake of measurement and methods.

These differences interact with each other, and collectively also exacerbate issues of distribution shift. Distribution shift is a technical machine learning problem that targets the differences in distribution between the training and test set, and can be motivated by fairness-related concerns. The methods proposed for distribution shifts are generally indifferent to which axis they target, working broadly for the domain of ``fairness.''
However, as the WILDS distribution shift leaderboard shows~\cite{koh2021wilds}, different models work better or worse for different domains. Some work targets subgroup population shift whereas other targets a broader domain shift; ultimately different methods will make different assumptions about the data. As prior work notes~\cite{yao2022ood}, unsuitable regularizers can have difficulty across diverse domains. One synthesizing work distinguishes between three forms of distribution shift, all of which are relevant to machine learning fairness: spurious correlations, low-data drift, and unseen data drift~\cite{wiles2022distshift}. After comparing 19 different methods, they find the results to be inconsistent over both different datasets as well as different attributes. In other words, just because something works well for race algorithmically, does not mean it necessarily will for gender. 
% The issues noted before from label collection again become relevant in terms of residual groups and within-group heterogeneity.  

This also means that toy experiments using pseudo-demographic groups, while certainly useful, may not necessarily generalize well to actual demographic groups. For example, in computer vision to get around the difficulties of using demographic attributes (e.g., studying gender on facial images might require inferring visual gender), black-and-white versus color or different colors are ways to create synthetic data biases~\cite{wang2020fairnessvision, kim2019colormnist}. These methods are analytically useful, but we should be cautious in over-indexing on their results. For example, in these synthetic datasets there are no residual groups that are neither colored nor black-and-white, and there is not ambiguity around what unifies the color group, as there is around what unifies the disability group. As Sophia Moreau writes ``When we try to test a theory of discrimination by appealing to happenings in fictitious societies... we bracket the complex social contexts in which real acts of discrimination occur. And these social contexts are, I shall argue, the key to understanding discrimination''~\cite{moreau2020faces}.

% Again, the fact that residual groups differ in how they are not accommodated by existing category schemas are relevant: the same method which adapts an algorithm suited for people of separate races to Multiracial, is likely not to be the same method which adapts an algorithm for men and women to Non-binary people.
\section{Case Study: Chatbot Math Tutoring}
Now that we have completed our presentation of axis-level and group-level differences, we present a brief case study to make concrete some of the considerations we have discussed so far.
Consider a fairness analysis of math tutoring in English through a chatbot for an introduction to algebra course. We may start by identifying which identity axes are relevant, and which harms are salient along each. For gender, we may decide the primary concerns are with respect to ensuring there is no differential treatment, and thus measure for invariance with respect to the student’s gender as well as the content of math word problems. While for race we similarly do not want students of different races to be treated differently, there are two correlated dimensions that could actually warrant differential treatment~\cite{wang2025diffaware}, in contrast to enforcing invariance: culture and linguistics. Prior work has shown that culturally and linguistically relevant word problems can make a difference for math education~\cite{driver2017wordproblemsolving}, so we should consider incorporating these findings for tutoring chatbots. Next, when considering age, we might expect many users to be adolescents. This brings up a level of school-appropriateness that would need to be enforced. However, there are also adult learners whom we would want to ensure not to infantilize. Finally, the most relevant axis here is likely learning ability. This is a core topic in education, and covered extensively in the literature~\cite{swanson2014handbook}. In our analysis, we may determine that axes like sexual orientation and marital status are not especially relevant. Taken together, the four axes of gender, race, age, and learning ability are not an exhaustive look at the fairness-related analyses that would need to be considered for this scenario, but an illustrative example of how axis specificity can shape the kinds of harms we measure and mitigate for. 
We should further consider intersectional interactions if we have specific hypotheses to support them. What we are proposing is in contrast to, for instance, measuring treatment or outcome disparities across all available groups. 

\section{Discussion}
In this work, we have argued for the specificity of identity axes and groups. This goes against the core values of machine learning, which strive for generality and plug-and-play methods~\cite{birhane2022values}.
% Consider how gender and sexuality studies is distinct from race studies, from disability studies. How African American studies is distinct from Latin American studies from Asian American studies. Of course there are similarities across these disciplines, and we certainly do not argue for specificity in all domains. 
Methods and measurements which generalize across axes help ease adoption, which otherwise serves as a serious roadblock for responsible machine learning. Theories which draw connections between different forms of discrimination can lead to grounded methods, and fruitful collaborations and coalitions across interest groups.
There truly are many parts of the machine learning pipeline that are amenable to the substitution of any axis. 
% And this ease of substitution causes increased adoption. 
For disabled or transgender groups whose algorithmic concerns have been historically ignored, something as small as adding a label can now allow the harm to those groups to be measured and potentially mitigated. However, what this framework misses is that specific groups can experience distinct harms. For example, misgendering is a unique harm towards transgender individuals that will be missed~\cite{ovalle2023whoiam, ovalle2024genderexclusive}.

Yet, even something as simple as measuring performance disparity between groups should acknowledge that identity axes are not wholly interchangeable. Measuring the accuracy difference between men and women isolates non-binary individuals, and needs to account for transgender individuals who may have different experiences from cisgender individuals. Whatever method is ultimately used for gender cannot be transported wholesale into measuring the accuracy difference between Black and White people in America, and accounting for all of the racial groups not included in this calculus. Counting Multiracial Americans is different than counting non-binary ones.

When we don't name which axes or groups we are working with, not only does it implicitly motivate creating methods which take into account the lowest common denominator among all the forms of oppression, but we keep in mind the ``norm'' groups: race with Black and White, gender with male and female. We miss all of the unique harms to the unnamed groups of disability, of non-binary people, of Indigenous people, of class differences.

One of the greatest fears by computer scientists of incorporating intersectionality into machine learning has been the problem of ``exploding groups'' where the number of groups to consider exponentially increases. However, incorporating context specificity actually helps to select which axes and groups are needed. In a given domain, one can consider historical context to understand which social groups have faced discrimination in the past, as well as consult existing regulations to identify legally impermissible practices and relevant protected groups. In other words, incorporating group specificity can \textit{reduce} the number of axes and groups studied for a particular harm. At the same time, in cases where greater numbers of groups is a ``benefit'' to the researcher or practitioner, such as being able to scale a benchmark to be much larger, the exploding groups problem has been leveraged in machine learning to artificially inflate the size of benchmarks. For example, a benchmark might claim over 450,000 unique sentence prompts. However, this scale is only achieved by having around 600 demographic groups and 26 sentence templates multiplied by a number of descriptor terms~\cite{smith2022holistic}. However, as we showed in Fig.~\ref{fig:examples}, this kind of generalization often does not make sense. In this work our goal is to confront fair machine learning's unyielding pursuit of generality. We do not wish to categorically stop this pursuit, but rather force justification in each instance as to whether generality makes sense. By doing so, we hope to broaden the scope of study to include those harms which are only relevant for one axis or group, treating them just as worthy of concern and science as those harms which are relevant to all axes.

\subsection*{Adverse Impact Statement}
% https://facctconference.org/2025/aguide
In engaging with work from non-computer science disciplines which have tended to be more abstract, we endeavored to bring in as much nuance as we can, while also not losing out on the concreteness and constructive recommendations that are favored in machine learning. In doing so, our translations of concepts may be imperfect and reformist, and may also become dated with time as we further develop our understanding of how best to operationalize different normative concepts. 
% However, in doing so we hope to demonstrate the feasability of engaging with this literature and bringing identity axis specificity into fair machine learning. 

\subsection*{Positionality}
I have lived and received my training in the United States. While the overarching argument of this piece applies globally, my background has influenced the regional emphasis of the examples included. 

\begin{acks}
I am grateful to Alida Babcock, Sunnie S. Y. Kim, Lizzie Kumar, Vikram Ramaswamy for feedback.
\end{acks}

\bibliographystyle{ACM-Reference-Format}
\bibliography{references}

\appendix

\end{document}